\documentstyle[12pt]{article}

\makeatletter

\newif\ifnoncomplete

\def\final{\noncompletefalse\typeout{** FINAL form (substyle:Sachiko)}}
\noncompletefalse

\newif\ifrefphysrev

\refphysrevfalse

\def \vol(#1,#2,#3){\ifrefphysrev{{\bf {#1}},
{#3} (19{#2})}\else{{{\bf {#1}}(19{#2}){#3}}}\fi}
\def \NP(#1,#2,#3){Nucl.\ Phys.\          \vol(#1,#2,#3)}
\def \PL(#1,#2,#3){Phys.\ Lett.\          \vol(#1,#2,#3)}
\def \PRL(#1,#2,#3){Phys.\ Rev.\ Lett.\   \vol(#1,#2,#3)}
\def \PRp(#1,#2,#3){Phys.\ Rep.\          \vol(#1,#2,#3)}
\def \PR(#1,#2,#3){Phys.\ Rev.\           \vol(#1,#2,#3)}
\def \PTP(#1,#2,#3){Prog.\ Theor.\ Phys.\ \vol(#1,#2,#3)}
\def \ibid(#1,#2,#3){{\it ibid.}\         \vol(#1,#2,#3)}

\def\thebibliography#1{
\section*{References\@mkboth
  {REFERENCES}{REFERENCES}}\list
  {[\arabic{enumi}]}{\setlength\labelwidth{2ex}
   \setlength\labelsep{0.05in} 
   \setlength\leftmargin{0.25in}
    \setlength\itemsep{0pt}
    \setlength\parsep{0pt}
   \itemsep\parskip
    \usecounter{enumi}}
    \def\newblock{\hskip .11em plus .33em minus -.22em}
    \sloppy
    \sfcode`\.=1000\relax}

\def\@bibitem#1{\item\if@filesw \immediate\write\@auxout
       {\string\bibcite{#1}{\the\c@enumi}}\fi\ignorespaces
       {\ifnoncomplete\reversemarginpar{\hspace*{-1.05in}\makebox[1in][l]
       {{\footnotesize{\sl [#1]}}}}\fi}%
       }

\def\@cite#1#2{\unskip\nobreak\relax
    {[#1]}} 

\def\citenum#1{{\def\@cite##1##2{##1}\cite{#1}}}
\def\citea#1{\@cite{#1}{}}

\newcount\@tempcntc
\def\@citex[#1]#2{\if@filesw\immediate\write\@auxout{\string\citation{#2}}\fi
  \@tempcnta\z@\@tempcntb\m@ne\def\@citea{}\@cite{\@for\@citeb:=#2\do
    {\@ifundefined
       {b@\@citeb}{\@citeo\@tempcntb\m@ne\@citea\def\@citea{,}{\bf ?}\@warning
       {Citation `\@citeb' on page \thepage \space undefined}}%
    {\setbox\z@\hbox{\global\@tempcntc0\csname b@\@citeb\endcsname\relax}%
     \ifnum\@tempcntc=\z@ \@citeo\@tempcntb\m@ne
       \@citea\def\@citea{,}\hbox{\csname b@\@citeb\endcsname}%
     \else
      \advance\@tempcntb\@ne
      \ifnum\@tempcntb=\@tempcntc
      \else\advance\@tempcntb\m@ne\@citeo
      \@tempcnta\@tempcntc\@tempcntb\@tempcntc\fi\fi}}\@citeo}{#1}}
\def\@citeo{\ifnum\@tempcnta>\@tempcntb\else\@citea\def\@citea{,}%
  \ifnum\@tempcnta=\@tempcntb\the\@tempcnta\else
   {\advance\@tempcnta\@ne\ifnum\@tempcnta=\@tempcntb \else \def\@citea{--}\fi
    \advance\@tempcnta\m@ne\the\@tempcnta\@citea\the\@tempcntb}\fi\fi}

\def\affiliation#1{\cr
\makebox[0in]{\parbox{8in}{\begin{center} {\sl #1}\end{center}}} \cr}
\def\@affiliation{}

\def\and{\cr \makebox[0in]{\rule[-1cm]{0mm}{1cm}and } \cr}

\def\maketitle{\par
 \begingroup
 \def\thefootnote{\fnsymbol{footnote}}
 \def\@makefnmark{\hbox
 to 0pt{$^{\@thefnmark}$\hss}}
 \if@twocolumn
 \twocolumn[\@maketitle]
 \else \newpage
 \global\@topnum\z@ \@maketitle \fi\thispagestyle{plain}\@thanks
 \endgroup
 \setcounter{footnote}{0}
 \let\maketitle\relax
 \let\@maketitle\relax
 \gdef\@thanks{}\gdef\@author{}\gdef\@title{}
 \gdef\@affiliation{} \let\affiliation\relax	%
 \let\thanks\relax}

\def\@maketitle{\newpage
 \null
 \vskip 0em plus 2em minus 0em     
 \ifx\@date\@empty\else
   \begin{flushright}
    {\ifnoncomplete(\today)
     \else{{\normalsize \@date}\\}\fi}      
   \end{flushright}
   \vskip 3em plus 2em minus 2em   
 \fi
 \begin{center}
  {\frtnsfb \@title \par}     
  \vskip 3em plus 1em minus 1.5em  
  {
   \lineskip .5em plus 0em minus .3em   
   \begin{tabular}[t]{c}\@author
   \end{tabular}\par}
\end{center}
 \par
 \vskip 6em plus 2em minus 4em}     

\def\abstract{\if@twocolumn
\section*{Abstract}
\else \normalsize
\fi}

\def\endabstract{\if@twocolumn\fi\par\clearpage}

\def\section{\@startsection {section}{1}{\z@}{3.5ex plus 1ex minus
    .2ex}{2.3ex plus .2ex}{\normalsize\bf}}
\def\subsection#1{\subsectioncom{\sc{#1}}}
\def\subsectioncom{\@startsection{subsection}{2}{\z@}
    {3.25ex plus 1ex minus .2ex}{1.5ex plus .2ex}{\small}}
\def\subsubsection{\@startsection{subsubsection}{3}{\z@}{3.25ex plus
1ex minus .2ex}{1.5ex plus .2ex}{\small}}

\def\@addmarginpar{\@next\@marbox\@currlist{\@cons\@freelist\@marbox
    \@cons\@freelist\@currbox}\@latexbug\@tempcnta\@ne
    \if@twocolumn
        \if@firstcolumn \@tempcnta\m@ne \fi
    \else
      \if@mparswitch
         \ifodd\c@page \else\@tempcnta\m@ne \fi
      \fi
      \if@reversemargin \@tempcnta -\@tempcnta \fi
    \fi
    \ifnum\@tempcnta <\z@  \global\setbox\@marbox\box\@currbox \fi
    \@tempdima\@mparbottom \advance\@tempdima -\@pageht
       \advance\@tempdima\ht\@marbox \ifdim\@tempdima >\z@
      \else\@tempdima\z@ \fi
    \global\@mparbottom\@pageht \global\advance\@mparbottom\@tempdima
       \global\advance\@mparbottom\dp\@marbox
       \global\advance\@mparbottom\marginparpush
    \advance\@tempdima -\ht\@marbox
    \global\ht\@marbox\z@ \global\dp\@marbox\z@
    \vskip -\@pagedp \vskip\@tempdima\nointerlineskip
    \hbox to\columnwidth
      {\ifnum \@tempcnta >\z@
          \hskip\columnwidth \hskip\marginparsep
        \else \hskip -\marginparsep \hskip -\marginparwidth \fi
       \box\@marbox \hss}
    \vskip -\@tempdima
    \nointerlineskip
    \hbox{\vrule \@height\z@ \@width\z@ \@depth\@pagedp}}

\def\ref#1{
    \@ifundefined{r@#1}{{#1}\@warning{Reference `#1'
    on page \thepage \space
    undefined}}{\edef\@tempa{\@nameuse{r@#1}}\expandafter
    \@car\@tempa \@nil\null}}
\def\refn#1{\@ifundefined{r@#1}{{#1}\@warning{Reference `#1'
    on page \thepage \space
    undefined}}{\edef\@tempa{\@nameuse{r@#1}}\expandafter
    \@car\@tempa \@nil\null}}

\def\endequationl{\eqno \@eqnnum 
$$\global\@ignoretrue}

\def\eqnarray{\stepcounter{equation}\let\@currentlabel=\theequation
\global\@eqnswtrue
\global\@eqcnt\z@\tabskip\@centering\let\\=\@eqncr
$$\arraycolsep\z@
\halign to \displaywidth\bgroup\@eqnsel\hskip\@centering
  $\displaystyle\tabskip\z@{##}$&\global\@eqcnt\@ne
  \hskip 2\arraycolsep \hfil$\displaystyle{{}##{}}$\hfil
  &\global\@eqcnt\tw@ \hskip 2\arraycolsep
  $\displaystyle\tabskip\z@{##}$\hfil
   \tabskip\@centering&\llap{##}\tabskip\z@\cr}
\def\mmodetrue{\mmode=\iftrue}

\def\eqnarrayl#1{\stepcounter{equation}\let\@currentlabel=\theequation
\label {#1}
\global\@eqnswtrue
\global\@eqcnt\z@\tabskip\@centering\let\\=\@eqncr
$$\arraycolsep\z@
\halign to \displaywidth\bgroup\@eqnsel\hskip\@centering
  $\displaystyle\tabskip\z@{##}$&\global\@eqcnt\@ne
  \hskip 2\arraycolsep \hfil$\displaystyle{{}##{}}$\hfil
  &\global\@eqcnt\tw@ \hskip 2\arraycolsep
  $\displaystyle\tabskip\z@{##}$\hfil
   \tabskip\@centering&\llap{##}\tabskip\z@\cr}
\def\label#1{
\@bsphack\if@filesw {
{\ifnoncomplete{\makebox[1in][r]{\footnotesize{\sl [#1]}}}\fi}%
\let\thepage\relax
   \xdef\@gtempa{\write\@auxout{\string
      \newlabel{#1}{{\@currentlabel}{\thepage}}}}
}\@gtempa
   \if@nobreak \ifvmode\nobreak\fi\fi\fi\@esphack}

\def\newlabel#1#2{
\@ifundefined{r@#1}{}{\@warning{Label `#1' multiply
   defined}}\global\@namedef{r@#1}{#2}}

\def\endeqnarrayl{\@@eqncr\egroup
      \global\advance\c@equation\m@ne$$\global\@ignoretrue}

\newif\if@numbersec \@numbersectrue
\def\appendix{\par\clearpage
  \setcounter{section}{0}
  \setcounter{subsection}{0}
  \def\thesection{\Alph{section}}
  \def\thesubsection{\arabic{subsection}}
  \@ifstar{\def\@sectname{Appendix}\@numbersecfalse}
          {\def\@sectname{Appendix~}\@numbersectrue}}

\def\thefigures#1{\par\clearpage\section*{Figures\@mkboth
  {FIGURES}{FIGURES}}\list
  {Fig.~\arabic{enumi}.}{\labelwidth\parindent\advance\labelwidth -\labelsep
      \leftmargin\parindent\usecounter{enumi}}}

\def\thetables#1{\par\clearpage\section*{Tables\@mkboth
  {TABLES}{TABLES}}\list
  {Table~\arabic{enumi}.}{\labelwidth-\labelsep
      \leftmargin0pt\usecounter{enumi}}}

\def\@sect#1#2#3#4#5#6[#7]#8{\ifnum #2>\c@secnumdepth
     \def\@svsec{}\else
     \refstepcounter{#1}\edef\@svsec{\ifnum #2=1 \@sectname
         \if@numbersec\csname the#1\endcsname\fi.\else
         \csname the#1\endcsname.\fi
        \hskip 1em }\fi
     \@tempskipa #5\relax
      \ifdim \@tempskipa>\z@
        \begingroup #6\relax
          \@hangfrom{\hskip #3\relax\@svsec}{\interlinepenalty \@M #8\par}
        \endgroup
       \csname #1mark\endcsname{#7}\addcontentsline
         {toc}{#1}{\ifnum #2>\c@secnumdepth \else
                      \protect\numberline{\csname the#1\endcsname}\fi
                    #7}\else
        \def\@svsechd{#6\hskip #3\@svsec #8\csname #1mark\endcsname
                      {#7}\addcontentsline
                           {toc}{#1}{\ifnum #2>\c@secnumdepth \else
                             \protect\numberline{\csname the#1\endcsname}\fi
                       #7}}\fi
     \@xsect{#5}}

\def\@sectname{}

 \def\thefootnote{\fnsymbol{footnote}}

\def \@magscale#1{ scaled \magstep #1}
\font\frtnsfb = cmssbx10 \@magscale2 
\newcount\ieq
\newcount\jeq
\newcount\keq
\def \eq{
\multiply\ieq by 2
\jeq=\ieq
\divide\jeq by 4
\multiply\jeq by 4
\ifnum\ieq=\jeq \end{eqnarray} \keq=1 
\else
\keq=2 \begin{eqnarray} \fi
\ieq=\keq
}

\def \mathbox(#1){\invisible\ifmmode{{#1}}\else{\mbox{${#1}$}}\fi}
\def \mbf(#1){\mbox{\boldmath{$#1$}}}
\def \abs(#1){\mathbox(\left|{#1}\right|)}
\def \bracket(#1){\mathbox(\left\langle{#1}\right\rangle)}
\def \brav(#1){\mathbox(\langle {#1}|)}
\def \cg(#1,#2,#3,#4,#5,#6){\mathbox({(#1\,#2\,#3\,#4|#5\,#6)})}
\def \comm(#1,#2){\mathbox(\left[{#1},{#2}\right])}
\def \dfdx(#1,#2){\mathbox(\frac{{\rm d}{#1}}{{\rm d}{#2}})}
\def \delfdelx(#1,#2){\mathbox(\frac{\partial{#1}}{\partial{#2}})}
\def \inprod(#1,#2){\mathbox({(#1\cdot #2)})}
\def \inprodij(#1){\mathbox({\inprod(#1_i,#1_j)})}
\def \intd(#1,#2){\mathbox({\int^#1_#2 \; \rmd})}
\def \eps(#1){\mathbox(\epsilon_{#1})}
\def \half(#1){\mathbox(\frac{#1}{2})}
\def \ketv(#1){\mathbox(|{#1}\rangle)}
\def \matele(#1,#2,#3){\mathbox(\left\langle {#1}|\,{#2}\,|{#3}\right\rangle)}
\def \mateled(#1,#2,#3){\mathbox(\left\langle
{#1}||\,{#2}\,||{#3}\right\rangle)}
\def \hatmbf(#1){\mathbox({\hat{\mbf({#1})}})}
\def \ninej(#1,#2,#3,#4,#5,#6,#7,#8,#9){\mathbox(\left\{\matrix
     {#1&#2&#3\cr#4&#5&#6\cr#7&#8&#9\cr}\right\})}
\def \rtov(#1,#2){\mathbox(\sqrt{{#1\over #2}})}
\def \sixj(#1,#2,#3,#4,#5,#6){\mathbox(\left\{\matrix
     {#1&#2&#3\cr#4&#5&#6\cr}\right\})}
\def \third(#1){\mathbox(\frac{#1}{3})}
\def \Trace(#1){\mathbox({\hbox{Tr} \left\{#1\right\}})}
\def \outprod(#1,#2){\mathbox({(#1\times #2)})}

\def \als{\mathbox(\alpha_s)}
\def \bra{\mathbox(\langle)}

\def \Del{\mathbox(\Delta)}
\def \etal{{\it et al.}}

\def \ie{{\it i.e.}}
\def \invisible{\mbox{$\rule{0mm}{1mm}$}}
\def \ket{\mbox{$\rangle$}}
\def \lamlam{\mbox{$(\lam_i\cdot\lam_j)$}}

\def \Lam{\mbox{$\Lambda$}}
\def \lam{\mbox{$\lambda$}}
\def \rmd{{\rm d}}

\def \Sig{\mbox{$\Sigma$}}

\def \vecsig{\mathbox({\vec \sigma})}

\def \vs{{\it vs.}}

\makeatother

\def\Del{\Delta}

\def\Lam{\Lambda}

\def\Sig{\Sigma}

\def\Voge{V_{\rm CMI}}
\def\als{\alpha_s}
\def\als{\alpha_s}
\def\etal{{\it et al.}}

\def\ie{{\it i.e.}}
\def\lamlam{\lam_i\cdot\lam_j}
\def\lam{\lambda}

\def\rv{\vec r}
\def\sigsig{\sigv_i\cdot\sigv_j}

\newcommand{\KY}{K.~Yazaki}
\newcommand{\MO}{M.~Oka}

\renewcommand{\bra}{\langle}
\renewcommand{\half}{\mbox{$\frac{1}{2}$}}
\renewcommand{\ket}{\rangle}

\renewcommand{\rtov}[2]{\mbox{$\sqrt{\frac{#1}{#2}}$}}

\renewcommand{\third}{\mbox{$\frac{1}{3}$}}
\def\Voge{\mathbox({V_{\rm OGE}})}
\def\Vcmi{\mathbox(V_{\rm CMI})}
\def\Vconf{\mathbox(V_{\rm CONF})}

\def\forth(#1){\mathbox(\frac{#1}{4})}
\def\lamlam{\lam_i\cdot\lam_j}

\def\sigsig{\vecsig_i\cdot\vecsig_j}

\begin{document}

\def\wave{\simeq}
\def\rtt{\sqrt{3}}
\def\DI#1{|\Delta I|=#1}

\final

\begin{flushright}
TIT/TP-239/NP\\
hepth@xxx/9310271\\
October 1993
\end{flushright}

\begin{center}
{\Large Hyperon-Nucleon Interaction in a Quark Model}
\footnote
{A lecture given at the International School Seminar on {\sl
Hadrons and Nuclei from QCD}, Tsuruga-Vladivostok-Sapporo,
August-September, 1993. }\\
\bigskip
MAKOTO OKA\footnote{e-mail: oka@phys.titech.ac.jp}\\
{\sl Department of Physics,
Tokyo Institute of Technology}\\
{\sl Meguro, Tokyo 152, Japan} \\
\bigskip
ABSTRACT
\end{center}

\begin{narrower}
A realistic hyperon ($Y$)-nucleon ($N$) interaction
based on the quark model and the one-boson-exchange potential
is constructed.
The Nijmegen potential model D with the SU(3) flavor symmetry is
modified with
a quark exchange interaction at the short-distance, which replaces
the short-range repulsive core in the original model.
The flavor-spin dependences of the short-range repulsion
are qualitatively different from the original hard-core potential.
We also study a two-body weak decay, $\Lambda N \to NN$, in the
quark model.  An effective weak interaction, where one-loop
QCD corrections are explicitly taken into account, is employed.
Differences from the conventional meson-exchange processes are discussed.
\end{narrower}

\bigskip
\section{Introduction}

In this lecture, I discuss two subjects:  First, I present our recent
attempt to constructing a realistic hyperon-nucleon interactions in
the quark model.  This part is based on the work done with Kenichiro
Ogawa and Sachiko Takeuchi{\cite{OSO}}.  The second part is devoted
for the study
of nonmesonic weak decays of $\Lambda$ in the direct quark processes,
which is done in collaboration with Takashi Inoue and Sachiko
Takeuchi{\cite{ISO}}.

\section{$Y-N$ interaction}

The short-distance repulsions between baryons seem universal for most
two-baryon interactions.
It is, for instance,  known from the study of hypernuclei that the
hyperon-nucleon interactions contain a short-range repulsion
similar to the nuclear force.
Why are the baryon-baryon interactions mostly repulsive at short
distances?   The simple quark model provides us with two possible
answers to this question.  Namely, there are two mechanisms
in the quark model which produce short-distance repulsion{\cite{OY80,Nagoya}}.

\medskip
The first (I) is due to the Pauli exclusion principle among the valence
quarks.  This can be most easily demonstrated in an extreme case,
$\Del^{++}(S_z={3\over 2})$ \vs\ $\Del^{++}(S_z={1\over 2})$, where
all the quarks are UP in the flavor and spin $\uparrow$ except for one
quark in the second $\Del$.  In a simple harmonic oscillator quark
model, if two baryons stick to each other with relative
0s harmonic oscillator state, one has to excite at least two of the UP
$\uparrow$ quarks to a higher single particle orbit in order to
satisfy the Pauli exclusion principle.
Mathematically, it means that the antisymmetrized product of two
$\Del^{++}$'s in relative 0s state vanishes:
\eq
   {\cal A}| \Del^{++}({3\over 2})\Del^{++}({1\over
        2})\chi_{0s}(R) \ket = |u^6, S_z=2, (0s)^6 \ket =0
\label{eq:FS}
\eq
where $\cal A$ is the quark antisymmetrization operator for all the six
quarks and $\chi_{0s}$ denotes the relative $\Del\Del$ wave function in
the harmonic oscillator 0s state.
Nonvanishing states require at least $2\hbar\omega$
excitation, which corresponds to the relative 1s state of two ground
state.   Namely, the relative 0s state of $\Del^{++}-\Del^{++}$ is
forbidden and all the allowed $\chi(R)$ has to be orthogonal to
$\chi_{0s}$.   Thus the relative wave function always has a node at
$R=R_c$, which is nearly independent of the energy and the
effective potential has a repulsive core of radius $R_c$.

One can interpret the above feature in a slightly different way.
By multiplying $ \bra{\Del^{++}\Del^{++}\delta(R-S)}|$
from the left of
eq(\ref{eq:FS}), one obtains
\eq
       \bra{\Del^{++}\Del^{++}\delta(R-S)} &&| {\cal A} |
             {\Del^{++}\Del^{++}\chi_{0s}(R)} \ket  \nonumber \\
    &=&    \int \bra{\Del^{++}\Del^{++}\delta(R-S)}| {\cal A}|
         {\Del^{++}\Del^{++}\delta(R-S')}\ket\,\chi_{0s}(S')\, dS' \nonumber\\
    &=&  \int N(S,S')\, \chi_{0s}(S')\, dS'
    =  e \, \chi_{0s}(S)
\eq
where $N(S,S')$ is the normalization integral kernel of the resonating
group method and $e$ is the eigenvalue of $N$ associated with
the eigenstate $\chi_{0s}$.  The forbidden state yields $e=0$, while
one obtains $e=1$ if no antisymmetrization is considered.  In general,
the eigenvalue $e$ gives a good indication of the ``forbiddeness'' of
the two-baryon system.  Namely,  if $e<1$, the channel
has a ``partially forbidden'' state and the baryonic potential has a
repulsion at short distances (or actually $R=0$).

When the same argument is applied to the hyperon-nucleon systems, one
finds that two $N\Sig$ channels, $N\Sig$ ($S=0$, $I={1\over2}$) and
$N\Sig$ ($S=1$, $I={3\over2}$), have small eigenvalues, $e=1/9$ and
$2/9$ respectively.  They thus have an almost forbidden state.  This
indicates a strong repulsion in the $S$wave $N\Sig$ interactions in
those channels{\cite{Theo1}}.

\medskip
The second mechanism (II) for the short range repulsion is driven by the
hyperfine interaction among quarks.  The success of the quark model
description of the meson-baryon spectrum owes largely to the spin-spin
interaction
\begin{equation}
   \Vcmi = - {\als\over 4} \sum_{i<j}
              {2\pi\over 3m_im_j} \, (\lamlam)  \, (\sigsig)\,
                  \delta(\rv_{ij}) \hfill
\label{eq:CMI}
\end{equation}
which is considered to come from the magnetic part of a gluon exchange
between quarks.
The importance of this interaction in the baryon spectrum is
manifested, for instance, in $N-\Del$, and
$\Lam-\Sig$ mass differences, and the negative
neutron mean charge square radius.

The importance of the hyperfine interaction in the
short-range $NN$ interaction has been pointed out in the quark
cluster model calculation\cite{OY80,Theo1}.
One finds that the spin-spin interaction
(\ref{eq:CMI}) produces a short-range repulsion not only for $NN$ but also
for other baryon-baryon interactions, such as $N\Lam$ and $N\Sig$.
Such calculations also indicate that the Pauli exclusion principle
(mechanism I) gives in general a
stronger short-range repulsion than the hyperfine interaction (II).

\section{Quark cluster model with the Nijmegen meson exchange potential}

We concentrate on the hyperon-nucleon ($YN$) interaction here and
present a realistic $YN$ interaction model, which incorporates
the quark exchange interaction at short distances and the meson
exchange potential at larger distances{\cite{OSO}}.
The antisymmetrization of six valence quarks with the one-gluon
exchange inteaction leads to a strong repulsion, whose range is
determined by the size of the baryon, and beyond its range
the conventional meson-exchange processes take
over and yield medium-long range attraction which binds nucleons
together into nuclei.

We follow the SU(3) symmetry for the meson-baryon couplings.  Indeed,
the $YN$ potential models, such as the Nijmegen models{\cite{Nij}} and
J\"ulich models{\cite{Julich}}, are based on the SU(3) symmetry.  In this
study, we employ the
meson-exchange part of the Nijmegen potential model D and instead of
using the hard cores in the original model, superpose it with the
quark exchange interaction at the short distance.

\medskip
The quark exchange interactions can be calculated in the quark cluster
model (QCM) approach{\cite{OY80}}.
We consider a valence quark model with a hamiltonian,
\eq    H=K+\Vconf +\Voge
\eq
where $K$ is the nonrelativistic quark kinetic energy term, $\Vconf$
stands for a quark confinement potential and $\Voge$ is the
Fermi-Breit potential for the one gluon exchange.
We employ the resonating group method (RGM) wave function for the
six-quark system, given by
\eq
    \Phi_{BB'} (1\sim 6) = {\cal A} [ \phi_B(1\sim 3)\,
\phi_{B'}(4\sim6) \,\chi(R)\,]
\eq
and the
integral equation, called the RGM equation, with kernels $H$
(Hamiltonian) and $N$ (Normalization):
\eq
  \int \left[ H(R,R') - E\, N(R,R') \right] \, \chi(R')\, dR' = 0
\label{eq:QCM}
\eq
are solved.
Nonlocality of the RGM equation comes from the antisymmetrization
of the quarks.
So far, we have not included effects of the instanton induced
interactions in this model{\cite{OT89,OT91}}.

\medskip
We introduce to the QCM equation
(\ref{eq:QCM})
the meson exchange potential,
which is borrowed from the Nijmegen model D in this study.
This can be done by adding
an integral kernel for the meson exchange potential, given by
\eq
       V(R,R') \equiv \int dR'' N^{1/2}(R,R'') V_{f} (R'')
      N^{1/2}(R'',R')
\eq
where $V_{f}$ is the Nijmegen meson exchange potential with the
appropriate form factor.  The form factor is chosen so as to be consistent
with the quark wave function of the baryon,
\eq
      V_{f} (R) \equiv \int \rho(x;R/2) V_N (x-y) \rho(y;-R/2)\, dx\,dy
\eq
where $V_N$ is the Nijmegen potential without the repulsive core and
the quark density of the baryon centered at $R/2$ is denoted by
$\rho(x;R/2)$.
In the QCM calculation, we employ the Gaussian wave function for the
quark for simplicity, and thus the corresponding form factor is given
also by a Gaussian.

We have five parameters in the model: the light quark mass
$m_q$, the ratio of the light and strange quark masses  $m_q/m_s$, the
strength of confinement $a$, the strength of the one-gluon exchange
potential $\alpha_s$, and
the size parameter $b$ for the Gaussian wave function of quarks in the
baryon.
In order to make the calculation consistent in kinematics, we choose
$m_q$ to be one-third of the average octet baryon mass, \ie, 383.7 MeV.
The ratio of the light/strange quark masses is fixed to 0.6, which
gives the $\Lam-\Sig$ mass difference.
The gluon coupling constant is chosen so as to reproduce the
$N-\Delta$ mass difference, and we also choose the
confinement $a$ so that the baryon state is stable against the breathing mode
excitation, \ie, $\partial E_B/\partial b = 0$.
The remaining parameter $b$ is sensitive to the $NN$
interaction, because it determines the size of the form factor and
also the range of the quark exchange interaction.  Therefore we leave
this as a free parameter and use the $NN$ scattering data to choose
the best value for $b$.
The QCM calculation with the Nijmegen D meson exchange potential can
fit the $NN$ $^1S_0$ scattering phase shift well for $b=0.56$ fm.
Then the other parameters are determined: $a=20.8$ MeV/fm, $\alpha_s=1.85$.

We calculate the scattering S matrices for various  $YN$ systems in
this model and
find that the qualitative predictions given above are
confirmed in the present model.
In Table 1, we summarize the properties of the short-distance
$YN$ interactions.
The lowest eigenvalue of the normalization kernel for each channel, given
in the Table,  distinguishes
the first (I) and the second (II) mechanisms for the short-range repulsion.
One sees that the Pauli exclusion principle gives a stronger repulsion
for the $N\Sig$ ($S=0$, $I={1\over2}$) and
$N\Sig$ ($S=1$, $I={3\over2}$) channels,
while the other channels show a mild repulsion which is generally
softer than the original Nijmegen model D.
The repulsions in these two $N\Sig$ channels are as strong as the Nijmegen
model F, which is known to provide not enough binding for $\Sig$ to
make a bound $\Sig$ hypernuclei.
Details of the model and the results will be published
elsewhere{\cite{OSO}}.

\renewcommand{\arraystretch}{1.5}
\begin{table}

\caption{
Eigenvalues of the normalization kernel and the effective core radii
for various S-wave $YN$ systems.
The ``type'' indicates the origin of the
repulsion, either  from the first (I) or the second (II)
mechanisms.
The effective core radii are obtained from the
scattering phase shifts in the present model.}

\begin{center}

\begin{tabular}{|rl|l|rl|}
\hline
  $BB'$         & ($J$,$I$) &  $e$   & type & effective core radius   \\
\hline
  $N   \Lam$    & (0,${1\over2}$)& 1           &  II  & 0.40 fm     \\
  $N   \Sig$    & (0,${1\over2}$)& {$1\over9$} &  I   & 0.68 fm     \\
  $N   \Lam$    & (1,${1\over2}$)& 1           &  II  & 0.34 fm     \\
  $N   \Sig$    & (1,${1\over2}$)& 1           &  II  & 0.30 fm     \\
\hline
  $N   \Sig$    & (0,${3\over2}$)& {$10\over9$} & II  & 0.48 fm     \\
  $N   \Sig$    & (1,${3\over2}$)& {$2\over9$} &  I   & 0.67 fm     \\
\hline
\end{tabular}
\end{center}
\end{table}

\section{Two-body Weak Decay of $\Lambda$}

The hyperon $\Lambda$ decays weakly into a nucleon and a pion in the
free space.
It, however, is suppressed in the nuclear medium by
the Pauli blocking on the final nucleon state, whose momentum is
less than 100 MeV/c for the $\Lambda$ decay at rest.  Indeed, in heavy
hypernuclei, the decay is predominantly the nonmesonic one, that is,
$\Lambda N \to NN$.  If we assume that the initial $\Lambda$ and the
nucleon are at rest, then the final relative momentum of $NN$ is
about 420 MeV/c and thus is well above the Fermi momentum.

Theoretical study of the $\Lambda N \to NN$ decay has traditionally
employed the meson ($\pi$, $\rho$, etc.) exchange mechanism, where one
of the meson-baryon verteces involves the weak transition
$s\to d${\cite{meson}}.
Contributions from the direct quark-quark weak interaction, $us\to ud$ or
$ds\to dd$,  have not been taken into account.  However, such direct
quark-quark processes may play significant roles, as the relative $NN$
momentum in the final state is not small.

Recent analyses of experimental data of decays of hypernuclei have
revealed some difficulties in the meson-exchange picture.  For instance,
the so-called $n-p$ ratio, i.e., the ratio $R_{np}$ of $\Lambda n \to
nn$ v.s. $\Lambda p \to np$ decay in the nucleus, is predicted very
small,  $R_{np}\wave 0.1$ in the meson-exchange picture.  This is due
to the strong contribution of the tensor force, which is preferred at the
large momentum transfer.  The tensor force selects the $S=1$, $I=0$
$pn$ final state and therefore $R_{np}$ becomes small.  The
experimental data, however, seem not to agree with the prediction, i.e.,
$R_{np} \wave 1$ in decays of light hypernuclei.
We argue that the direct quark process, which does not follow the
$I=0$ selection rule, may enhance the $n-p$ ratio.

The mesonic weak decays of hyperons have been tested for the $|\Delta
I|= {1\over 2}$ rule and are known to satisfy the rule to about 5\%
error.   The
same rule for the nonmesonic weak processes, like $\Lambda N \to NN$,
is not confirmed yet.  Indeed, an analysis of the decay of the  $A=3$ and 4
hypernuclei claims that the $|\Delta I|= {1\over 2}$ rule is not
satisfied{\cite{Shoe}}.
It is therefore urgent to clarify the mechanism of the
$|\Delta I|= {1\over 2}$ rule in the free hyperon decays and to study
whether the same mechanism restricts the nonmesonic decays to $|\Delta
I|= {1\over 2}$ as well.

In the study of the meson-exchange processes, the $|\Delta I|= {1\over
2}$ rule is assumed from the beginning, implemented in the $\Lambda\to
N\pi$ vertex.  We instead employ the effective quark-quark weak
hamiltonian, which contains both the $|\Delta I|= {1\over 2}$ and
$|\Delta I|= {3\over 2}$ components.  Although the $|\Delta I|=
{3\over 2}$ part has a small overall coefficient, we will see that the
matrix elements for the $\Lambda N\to NN$ decay may not be small
compared to the $|\Delta I|= {1\over 2}$ component.

\bigskip
The effective weak hamiltonian describing $\Delta S = \pm 1$ processes,
given by several authors{\cite{VSZ,Paschos}}, is
\eq
  H_{eff}^{\Delta S=1}\left(Q^2\sim{\mu}^2\right)=
  -\frac{G_f}{\sqrt 2}\sum_{r=1,r\ne 4}^6K_rO_r
\eq
where the four-quark operators, $O_k$ ($k= 1$, 2, 3, 5 and 6) are
defined by{\cite{VSZ}}
\eq
 O_1 &=& (\bar d_{\alpha}s_{\alpha})_{V-A}
          (\bar u_{\beta}u_{\beta})_{V-A}
         -(\bar u_{\alpha}s_{\alpha})_{V-A}
           (\bar d_{\beta}u_{\beta})_{V-A}\\
  O_2 &=& (\bar d_{\alpha}s_{\alpha})_{V-A}
           (\bar u_{\beta}u_{\beta})_{V-A}
         +(\bar u_{\alpha}s_{\alpha})_{V-A}
           (\bar d_{\beta}u_{\beta})_{V-A}\nonumber\\
     &+& 2(\bar d_{\alpha}s_{\alpha})_{V-A}
         (\bar d_{\beta}d_{\beta})_{V-A}
        +2(\bar d_{\alpha}s_{\alpha})_{V-A}
         (\bar s_{\beta}s_{\beta})_{V-A}\\
  O_3 &=& 2(\bar d_{\alpha}s_{\alpha})_{V-A}
           (\bar u_{\beta}u_{\beta})_{V-A}
         +2(\bar u_{\alpha}s_{\alpha})_{V-A}
           (\bar d_{\beta}u_{\beta})_{V-A}\nonumber\\
     &-& (\bar d_{\alpha}s_{\alpha})_{V-A}
         (\bar d_{\beta}d_{\beta})_{V-A}
        -(\bar d_{\alpha}s_{\alpha})_{V-A}
         (\bar s_{\beta}s_{\beta})_{V-A}\nonumber\\
     &=&  {1\over3} O_2 + O_3(\Delta I={3\over 2})\\
  O_5 &=& (\bar d_{\alpha}s_{\alpha})_{V-A}
       (\bar u_{\beta}u_{\beta}+\bar d_{\beta}d_{\beta}
      + \bar s_{\beta}s_{\beta})_{V+A}\\
  O_6 &=& (\bar d_{\alpha}s_{\beta})_{V-A}
       (\bar u_{\beta}u_{\alpha}+\bar d_{\beta}d_{\alpha}
      + \bar s_{\beta}s_{\alpha})_{V+A}
\eq

The coefficients (Table 2) for the above six four-quark operators are
calculated by using the renormalization group technique within
the one-loop QCD corrections included{\cite{Paschos}}.
The most prominent feature of this effective hamiltonian is that the
QCD correction enhances the $O_1$ component while the other terms are
suppressed. This is the main mechanism for the $\Delta I= 1/2$ enhancement.

\begin{table}
\caption{Strengths of the weak effective four-fermi verteces}
\begin{center}

\begin{tabular}{ccccc}
 $K_1$ & $K_2$ & $K_3$ & $K_5$ & $K_6$\\ \hline
 $-0.265$ & 0.010  & 0.026 & 0.003 & $-0.020$
\end{tabular}
\end{center}
\end{table}

This effective hamiltonian has been used for the calculations of the
nonleptonic decay of strange mesons and baryons.{\cite{VSZ,Paschos,Fujii}}
It is found that although the $\Delta I= 1/2$ enhancement is
significant in those  decays,  agreement to experiment is
not always achieved quantitatively.  Some suggest that the decay
amplitudes are sensitive to the meson and
baryon wave functions.

We will adopt the above effective hamiltonian, but also consider the
case where the $\Delta I= 3/2$ component of the operator $O_3$ is
omitted and compare the results in order to see the effect of the
$\Delta I= 3/2$ contribution.

In calculating the decay amplitude for $\Lambda N \to NN$, we employ
the quark cluster picture for the two baryon systems.  In the present
calculation{\cite{ISO}}, we choose the most simple wave functions for
the initial
and the final states.  First, we assume that the baryon consists of
three valence quarks, whose orbital wave function is a harmonic
oscillator eigenstate.
\eq
\phi(1,2,3)^{\mbox{orb}}=
\left(\frac{1}{2\pi b^2}\right)^{\frac34}
   \left(\frac{2}{3\pi b^2}\right)^{\frac34}
    \mbox{exp}\left\{-\frac{1}{4b^2}{\vec{\xi}_{12}}^2\right\}
   \mbox{exp}\left\{-\frac{1}{3b^2}{\vec{\xi}_{12-3}}^2\right\}
\eq
The six quark wave function is given by
\eq
\vert \Lambda N\rangle
                      &= &{\cal A}^6\vert \phi(1,2,3)\phi(4,5,6)
                          \chi_0(\vec{R})\rangle\\
\vert NN\rangle
               &=&{\cal A}^6\vert \phi(1,2,3)\phi(4,5,6)
               \chi(\vec{R}')\rangle
\eq
where ${\cal A}^6$ is the antisymmetrization operator for six quarks, and
$\phi$ is the internal wave function of the baryon.
The flavor-spin wave function of the baryon is taken to be purely the SU(6)
wave functions, which is known to be a good approximation.

In this article, we only consider the simplest case,
 in which the initial $\Lambda$ and $N$
are on top of each other and therefore the orbital part of the initial
wave function is that of the $(0s)^6$ configuration in the harmonic
oscillator shell model.  Similarly, the final state is assumed simply
a plane wave of two nucleon clusters.
\eq
\chi_0(\vec{R})&=&\left(\frac{3}{2\pi b^2}\right)^{\frac34}
          \mbox{exp}\left\{-\frac{3}{4b^2}\vec{R}^2\right\} \\
\chi(\vec{R}') &=&\left(\frac{1}{2\pi}\right)^{\frac32}
          \mbox{exp}\left\{i\vec{k}\cdot\vec{R}'\right\}
\eq
Although these choices of the wave functions are not realistic,
they will clarify the qualitative difference between
the meson-exchange and the direct-quark processes, which is the
purpose of this preliminary study.  A full-range
calculation using the realistic two-baryon wave functions is underway.

Because we employ the nonrelativistic valence quark picture for the
wave functions, the effective hamiltonian is also approximated by
adopting the Breit-Fermi nonrelativistic expansion up to $1/c$.

\begin{table}
\caption{Possible initial and final quantum numbers
for the initial $L=0$ transition and the calculated transition
invariant matrix elements in $10^{-9}$ MeV$^{-1/2}$.}
\begin{center}

\begin{tabular}{cccccc}
\mbox{channel}&  \mbox{isospin}  &  \mbox{spin--orbital} &
&  full
       &  $\Delta I=\frac32$  omitted\\
\noalign{\hrule}
1 &   $p\Lambda\to pn$   &   $^{1}S_0\to{}^{1}S_0$  & $a_p$ & 2.68  & 5.52\\
2 &                                  &
     $^{1}S_0\to {}^{3}P_0$  & $b_p$ & 2.07 & 0.48\\
3 &                                  &
      $^{3}S_1\to {}^{3}S_1$  & $c_p$  & 6.67 & 6.67 \\
4 &                                  &
     $^{3}S_1\to {}^{1}P_1$  & $e_p$ & $-0.39$ & $-0.39$ \\
5 &                                  &
     $^{3}S_1\to {}^{3}P_1$  & $f_p$ & $-1.31$ & $-1.22$ \\
\noalign{\hrule}
6 & $n\Lambda\to nn$     &   $^{1}S_0\to{}^{1}S_0$ & $a_n$  & 9.80 & 7.80 \\
7 &                                  &
      $^{1}S_0\to {}^{3}P_0$ & $b_n$ & $-0.45$ & 0.68  \\
8 &               &
      $^{3}S_1\to {}^{3}P_1$ & $f_n$ & $-1.66$ & $-1.72$ \\
\noalign{\hrule}
\end{tabular}
\end{center}
\end{table}

\medskip
If we restrict our initial state to $L=0$,  there exist eight possible
combinations, given in Table 3, of $L$, $S$, and $J$  for the
initial and final states.
We note that the $I=1$ final states are allowed both for
($\Lambda n \to nn$) and ($\Lambda p \to pn$), while the $I=0$
states are not possible for ($\Lambda n \to nn$).
Thus  we have
5 ($\Lambda p \to pn$) and 3 ($\Lambda n \to nn$) matrix elements,
which are labeled from $a$ through $f$ in Table 3, according to a
widely used notation{\cite{BD}}.
The results of the calculation are also given in Table 3.
Eight amplitudes give all the information for the $\Lambda N\to NN$ weak
decay in the present calculation.   One sees that the parity conserving matrix
elements ($a$ and $c$) are dominant both in $I=0$ and $I=1$ channels.
The last column of Table 3 shows
the results
after omitting  the
$|\Delta I|= {3\over 2}$ components of the $O_3$ operator.
We find a significant
contribution of the $|\Delta I|= {3\over 2}$ matrix elements.

\begin{table}
\caption{Calculated observables}
\begin{center}

\begin{tabular}{ccc}
    & $\mbox{full}$ & $\Delta I=\frac32\mbox{omitted}$ \\
\noalign{\hrule}
$\Gamma_p$ ($10^{8}\mbox{sec}^{-1}$)&  2.82  &  3.16  \\
$\Gamma_n$ ($10^{8}\mbox{sec}^{-1}$)&  1.96  &  1.31  \\
\noalign{\hrule}
$R_{np}$ &  0.70  &  0.42\\
$\eta_p$ & 0.070 & 0.031\\
$\eta_n$ & 0.088 & 0.15\\
\noalign{\hrule}
$a_1(p)$ & $-0.28$  & $-0.24$\\
$a_1(n)$ &  0 & 0\\
\noalign{\hrule}
\end{tabular}
\end{center}
\end{table}

In Table 4, we summarize the calculated decay rates with the initial
spin averaged and the final states summed up.
The $n-p$ ratio,
$R_{np} \equiv \Gamma_n/\Gamma_p$,
the ratio of the parity
violating (PV)  v.s. the parity conserving (PC) contributions,
$\eta$,
and the decay asymmetry parameter, $a_1$, for
the $\Lambda p \to pn$ and the $\Lambda n\to nn$ decays
are also given in Table 4.
We find that the $n-p$ ratio in the
present calculation is much larger than that obtained in the
meson-exchange calculation.
It is very encouraging.  Although the present calculation assumes a
very naive wave function for the initial and final states, one sees at
least the direct-quark process has the right direction to improve the
meson exchange result, which is too small to account for the experimental
value.

The decay asymmetry parameter describes the angular distribution of
the outgoing two nucleons in the rest frame,
\eq
	W(\theta) = 1 + a_1 \, {\cal P}_{\Lambda} \, P_1(\cos\theta)
\eq
where ${\cal P}_{\Lambda}$ is the polarization of the lambda particle
in the nucleus.
Recent experiment done at KEK indicates a large $a_1$ for light
hypernuclei.  The data is consistent with $a_1 \wave -1.0 \pm 0.4$.
Our calculation yields the correct sign, but the magnitude seems too
small.

\section{Conclusion and Discussion}

We present a quark model analyses of the hyperon-nucleon systems both
for the strong and weak interaction processes.
The realistic strong $YN$ interactions, which are nonlocal due to the
quark antisymmetrization effects, are proposed using the quark cluster
model approach with the Nijmegen model D meson exchange potential.
The main difference between the original Nijmegen model and our
interaction arises in the spin-isospin dependence of the $YN$ short
range interactions.  Especially, $\Sig N$ with $S=0$, $I=1/2$ and
$S=1$ and $I=3/2$ have strong repulsion at the short distance in the
quark model and may make the bound $\Sig$ hypernuclei unplausible.

We also present a  quark model calculation of the direct-quark
processes for the weak $\Lambda N\to NN$ decay, which can be observed
exclusively in decays of hypernuclei.  Assuming simple initial and
final wave functions, we find that the calculated
decay rates are comparable to the meson exchange contributions in
magnitudes and show qualitatively distictive properties.
This is encouraging because the direct-quark processes may resolve the
discrepancies between experiment and the calculated results in the
meson-exchange mechanism.
The further study of the weak process is underway{\cite{ISO}}.

\end{document}